\documentstyle[12pt,aaspp4]{article}
\begin{document}
\parindent=1.0cm

\title{THE BRIGHTEST AGB STARS IN THE INNER BULGE OF M31}

\author{T. J. Davidge \altaffilmark{1}}

\affil{Canadian Gemini Office, Herzberg Institute of Astrophysics,
\\ National Research Council of Canada, 5071 W. Saanich Road, \\ Victoria,
British Columbia, Canada V9E 2E7 \\ {\it email:tim.davidge@nrc.ca}}

\altaffiltext{1}{Visiting Astronomer, Canada-France-Hawaii Telescope, which is 
operated by the National Research Council of Canada, the Centre National del la
Recherche Scientifique, and the University of Hawaii}

\begin{abstract}

	$JHK$ images with angular resolutions approaching the diffraction 
limit of the 3.6 meter Canada-France-Hawaii Telescope are used to investigate 
the bright AGB content of the M31 bulge. The AGB-tip in a field 
2.6 arcmin from the galaxy center occurs at $K = 15.6$, which 
is significantly fainter than measured in previous ground-based 
studies that sampled similar projected distances from the center of M31, but 
were affected by crowding. Within 2.6 arcmin of the center of M31 the 
number density of bright AGB stars scales with $r$-band surface brightness, and 
the $K$ brightness of the AGB-tip does not vary measureably with radius. 
It is concluded that the infrared bright AGB stars 
(1) belong to the bulge, and not the disk, and (2) are well mixed throughout 
the inner bulge, suggesting that they formed at a time when the overall 
structural properties of the M31 bulge were imprinted. The bolometric 
luminosity functions (LFs) of the M31 bulge and Baade's Window 
are in excellent agreement, while the brightest AGB stars in the M31 bulge, 
the Galactic bulge, and M32 have similar M$_K$. Barring a fortuitous tuning of 
age and metallicity to produce AGB-tips with similar brightnesses, it is 
suggested that the brightest stars in M32 and the bulges of M31 and the 
Milky-Way belong to an old, metal-rich population; these stars are bright not 
because they have a young or intermediate age, but because they have a high 
metallicity.

\end{abstract}


\section{INTRODUCTION}

	As the nearest large external galaxy, M31 provides an unprecedented 
laboratory for probing galaxy evolution. The stellar content in the central 
few arc minutes of M31 provides a fossil record that can be used to trace the 
evolution of the bulge, and possibly even the origins of the super-massive 
black hole (Dressler \& Richstone 1988; Kormendy 1988) and double nucleus 
(Lauer et al. 1993). Because of the obvious problems posed 
by high stellar densities, most studies of the 
stellar content near the center of M31 have relied upon integrated spectra. 
Bica, Alloin, \& Schmidt (1990) used evolutionary synthesis techniques to 
conclude that the inner bulge of M31 is dominated by an old, metal-rich 
component, although signatures of an intermediate-age population 
contributing 10 -- 20\% of the visible light, and even a 
component with an age of a few Myr, were also found. 
Davidge (1997) used a grid of long-slit spectra to map the 
stellar content within 30 arcsec of the nucleus, and concluded 
that (1) younger, more metal-rich populations occur at smaller radii, (2) 
[Mg/Fe] varies with radius, in the sense that [Mg/Fe] decreases towards 
smaller radii, as expected if the youngest stars formed from 
material enriched by SN I, whereas the main body of the bulge 
formed from material enriched by SN II, and (3) the radial gradients in 
stellar content differ from those in the elliptical galaxies 
NGC 3379 and NGC 4472, where age and [Mg/Fe] appear not to change with radius. 
Sil'Chenko, Burenkov, \& Vlasyuk (1998) conclude that the stars near the 
nucleus of M31 are three times younger than those in the surrounding bulge. 

	Studies of resolved stars provide a direct means of probing stellar 
content and checking the results from integrated spectra, 
although only the brightest objects can typically 
be observed, and data with high angular resolutions are required to 
overcome crowding. There is an absence of main sequence stars hotter 
than B0V (Bohlin et al. 1985) in the M31 bulge, although a large population of 
relatively faint hot stars is present in the UV images (Brown et al. 1998). 
While originally thought to be post-AGB objects (King et al. 1992, Ferguson 
\& Davidsen 1993), it is more likely that these sources are 
evolving on the horizontal branch (Brown et al. 1998). 
Rich \& Mould (1991), Davies, Frogel, \& Terndrup (1991), Rich, Mould, \& 
Graham (1993), Rich \& Mighell (1995) Davidge et al. (1997), and Stephens et 
al. (2001a, b) studied the near-infrared properties of stars in various M31 
bulge fields. Rich et al. (1989) find that the brightest giants in the M31 and 
Galactic bulges are spectroscopically similar, while Rich \& Mould 
(1991) and Rich, Mould, \& Graham (1993) conclude that the M31 bulge contains 
stars that are more luminous than the brightest 
stars in the Galactic bulge. Using $J, H$, and $K$ observations obtained 
with NICMOS, Stephens et al. (20001b) investigated the stellar content in 
fields surrounding five metal-rich M31 globular clusters, and concluded 
that (1) the LF of the M31 bulge at near-infrared wavelengths is similar 
to that in Baade's Window (BW), and (2) there is no evidence for a significant 
population of bright young stars. The existence of a very bright 
stellar component in the M31 bulge has long been questioned, and it has been 
argued that such a population could be an artifact of crowding (DePoy et al. 
1993; Renzini 1998; Jablonka et al. 1999) or, at larger 
distances from the nucleus, contamination from disk stars (Davies et al. 1991), 
although Rich et al. (1993) conclude that disk contamination is not an issue 
for fields within 500 parcsecs of the galaxy center.

	In the present paper high angular resolution $J, H,$ and $Ks$ images 
are used to investigate the brightness and spatial distribution of stars 
evolving on the upper portions of the AGB in two M31 inner bulge fields, with 
the goal of gaining insight into the nature and origin of the brightest stars 
in the M31 bulge. The paper is structured as follows. 
The observations and procedures used to reduce the data, as well as the 
methods used to measure stellar brightnesses, are discussed in \S 2. The 
stellar contents of the two fields are investigated in \S 3 and \S 4, while a 
summary and discussion of the results follows in \S 5. 

\section{OBSERVATIONS, DATA REDUCTION, \& PHOTOMETRIC MEASUREMENTS}

	The data were recorded during two observing runs with the KIR imager 
and CFHT Adaptive Optics Bonnette (AOB; Rigaut et al. 1998). KIR contains a 
$1024 \times 1024$ Hg:Cd:Te detector, with each pixel subtending 0.034 arcsec 
on a side; hence, the total imaged field is $35 \times 35$ arcsec. The CFHT AOB 
uses natural guide stars as reference beacons, and contains 
a 19 element curvature wavefront sensor (WFS) and a 19 electrode 
deformable bimorph mirror. The signal from the WFS is sampled at a frequency of 
1000 Hz. The system is operated with automatic mode gain optimization to adapt 
to guide star brightness and atmospheric conditions, and is designed to deliver 
diffraction-limited angular resolution at near-infrared wavelengths during 
median Mauna Kea seeing conditions; however, if the seeing 
degrades from its median value, or the guide star is faint, then the 
delivered FWHM is intermediate between the diffraction-limited and uncorrected 
cases. 

	A field 2.6 arcmin south west of the galaxy nucleus, centered on 
the $R = 13$ star GSC 02801--02015 (RA 00:42:45.1, DEC 
$+$41:13:31.3 E2000), which served as the reference source for AO compensation, 
was observed through $J, H$, and $Ks$ filters on the 
night of UT September 5 1998. The total integration time was 20 minutes per 
filter, with five 60 sec exposures being recorded at each corner of a 
$0.5 \times 0.5$ arcsec square dither pattern. 
The seeing conditions were poor by Mauna Kea standards, and stars in the final 
images have very low Strehl rations, with FWHM = 0.35 arcsec. This 
field will be refered to as the `bulge' field for the remainder of the paper.

	The central field of M31 was observed through $J, H,$ and $Ks$ filters 
during the night of UT September 11, 2000, with the semi-stellar nucleus of 
M31 serving as the reference source for AO compensation. The exposure times 
and observing strategy were the same as those employed for the bulge field. 
The seeing conditions were much better than during the 1998 run, and stars in 
the final images have FWHM = 0.27 arcsec in $J$, 0.15 arcsec in $H$, and $0.17$ 
arcsec in $Ks$; the diffraction pattern produced by the telescope optics is 
clearly evident in the point spread functions (PSFs) constructed from the $H$ 
and $Ks$ images. This field will be refered to as the `central' field for the 
remainder of the paper.

	The data were reduced using the procedures described by Davidge \& 
Courteau (1999), which correct for dark current, flat-field 
variations, thermal emission along the optical path, and variability in the 
DC sky level. The final $Ks$ images of both fields are shown in Figures 1 and 2.

	The brightnesses of individual stars were measured with the PSF-fitting 
program ALLSTAR (Stetson \& Harris 1988), using PSFs and star lists obtained 
from tasks in DAOPHOT (Stetson 1987). A single PSF was constructed for each 
image. While anisoplanicity causes the PSF to vary with distance from the 
reference source, past experience indicates that the use of a single PSF does 
not introduce large photometric uncertainties over the KIR field (e.g. Davidge 
\& Courteau 1999; Davidge 2001). Aperture corrections were determined from 
growth curve analysis of the PSF stars after subtracting neighboring objects, 
and the estimated uncertainty in the aperture 
corrections is $\pm 0.05$ magnitude. Both datasets were 
recorded during clear sky conditions, and the photometric
calibration was determined from observations of UKIRT faint standard stars 
(Casali \& Hawarden 1992). The brightnesses of the standard stars were 
measured using apertures established by growth curve analysis, and the 
uncertainties in the photometric zeropoints is $\pm 0.02 - 0.03$ mag.

	Unresolved stars create a non-uniform 
background in both fields, which complicates efforts to measure local sky 
levels and obtain reliable photometry. This background was removed using 
the iterative technique described by Davidge, Le F\'{e}vre, \& Clark (1991). 
Background structure was mapped by applying a $2 \times 2$ arcsec running 
median filter to star-subtracted frames, and the result was subtracted from the 
images prior to obtaining a final set of photometric measurements. 
This procedure does not contribute additional random noise to the data 
because of the large smoothing kernel used to construct the background frame.

	Artificial star experiments were used to estimate the 
uncertainties in the photometric measurements, calibrate 
systematic effects, which can be significant for stars near the faint limit 
of the data, and estimate incompleteness. 
The brightnesses of artificial stars were measured using the same 
procedures that were applied to the actual data, including background 
subtraction. The artificial star experiments assume a single PSF for each 
field, and so do not account for anisoplanicity; however, the good 
agreement between the predicted and observed scatter in the bulge field CMDs 
(\S 3) confirms that anisoplanicity does not dominate the uncertainties 
in the photometry. 
 
\section{THE BULGE FIELD}

\subsection{The CMDs and Comparisons with Baade's Window}

	The $(K, H-K)$ and $(K, J-K)$ CMDs of the bulge field are shown in 
Figure 3. The uncertainties predicted from the 
artificial star experiments match the scatter 
in the data. Not only does this agreement lend confidence to the predictions 
made from the artificial star experiments, but it also indicates that random 
uncertainties in the photometric measurements, rather than star-to-star 
differences in intrinsic properties, are the main sources of scatter in these 
data. 

	The stars plotted in Figure 3 are evolving on the upper AGB. 
There is a locus of stars near the blue edge of the $(K, J-K)$ CMD 
that runs from $J-K = 1.4$ at $K =16.4$ to $J-K = 1.6$ at $K = 15.6$; 
the $(K, H-K)$ and $(K, J-K)$ CMDs of the M32 outer field 
studied by Davidge (2000a) also show a well-defined sequence at the bright end.
This sequence disappears near $K = 16.4$ in Figure 3, at which point the CMD 
broadens. This broadening is not due to the 
RGB-tip, as this feature occurs near $K = 18$ at the distance of M31.

	The projected distance of the bulge field from the nucleus of M31 is 
0.6 kpc, which is comparable to the projected separation between BW 
and the Galactic Center (GC). Frogel \& Whitford (1987) studied the 
photometric properties of M giants in BW, and found that late M giants 
typically have $J-K \sim 1.3$ and $H-K \sim 0.4$. If shifted to the distance of 
M31, a late M giant from BW would have $K \sim 17$, and 
the colors of stars in the M31 bulge field at this brightness match the 
typical values for M giants in BW.

	The peak brightness in the M31 bulge field 
is similar to that in BW. The brightest star in Figure 3 has 
$K = 15.4$, and is 0.3 mag brighter than the main body of the giant branch, 
which has M$_K = -8.7 \pm 0.1$ if $\mu_0 = 24.4$ (van den Bergh 2000); this 
distance modulus for M31 is adopted for the remainder of the paper. 
The brightest star in the Frogel \& Whitford (1987) sample 
is number 239, which has M$_K = -9.5$, while 
star number 181, which is their second brightest object, has 
M$_K = -9$. The 0.5 mag gap in $K$ between the brightest and second brightest 
star in the Frogel \& Whitford compilation suggests that objects like star 239 
may be rare, perhaps because they are long period variables viewed at the peak 
of their cycle. Consequently, if star 181 is adopted as a representative 
example of the bright stellar content in BW, then the peak $K$ 
brightness in the M31 bulge field is similar to that in BW.

	Frogel \& Whitford (1987) derived $K-$band bolometric corrections 
for giants in BW as a function of $J-K$, and their calibration was applied to 
the M31 bulge field measurements to investigate the bolometric LF of stars in 
this field. The resulting LF, corrected for incompleteness using the 
results from the artificial star experiments, is shown in 
Figure 4. Also shown in this figure is the bolometric LF of BW, 
constructed from the entries listed in Table 1 of Frogel \& Whitford (1987), 
but assuming a distance modulus of 14.5 (Reid 1993). 
The BW LF, which has been scaled to match the total number of stars 
in the M31 LF when M$_{bol} < -4$, agrees with the M31 measurements 
to within the estimated uncertainties, indicating that the bright stellar 
contents in the bulges of M31 and the Milky-Way are similar.

\subsection{Comparisons with Previous Near-Infrared Studies of the M31 Bulge}

	Rich \& Mould (1991) and Rich et al. (1993) obtained deep 
$J$ and $K$ images of five M31 bulge and inner disk fields, and 
the ridgelines in the CMDs constructed from these data have $J-K$ 
colors that are consistent with the CFHT bulge field data. The Rich et al. 
data also show a tendency for peak brightness to 
increase towards smaller radii. For example, the brightest giants in Rich et 
al. (1993) Field 1, located 2.1 arcmin from the center of M31, have $K = 
14.5$, while the brightest stars in their Field 3, located 3.9 arcmin from 
the nucleus, have $K = 15$; the main body of the giant branch in their Field 3 
starts at $K = 15.5$, which is in rough agreement with what is seen in Figure 3.

	The CFHT bulge field and Rich et al. (1993) Field 1 have similar 
projected distances from the center of M31, and the peak brightnesses 
measured from these data differ by roughly 1 magnitude in $K$, in the sense 
that the peak brightness infered from the Rich et al. data is brighter. 
The 0.35 arcsec FWHM angular resolution of the CFHT 
data is much better than the Rich et al. (1993) data, which have FWHM in the 
range 1 -- 1.2 arcsec, and this can have a major 
impact on the measured peak brightness; indeed, Rich et al. (1993) 
suggest that the peak brightness in their Field 1 may be affected 
by crowding. To confirm this, the CFHT data were 
convolved with a Gaussian to simulate 1 arcsec seeing. DAOPHOT and ALLSTAR 
were then used to measure stellar brightnesses in the resulting smoothed 
image, and the $(K, J-K)$ CMDs of the raw and smoothed datasets are compared 
in Figure 5. The peak stellar brightness in the smoothed dataset is markedly 
brighter than in the unsmoothed data, confirming the suggestion made by 
Rich et al. (1993) that the brightest stars in their Field 1 are blends.

	Stephens et al. (2001b) investigated the stellar content of 
fields surrounding the globular clusters G174 and G177, which have projected 
distances from the center of M31 that are comparable to the CFHT bulge field. 
These data reveal unexpected field-to-field differences in 
stellar density, as the density of stars is greatest near 
G177, even though G174 is significantly closer to the center of M31. The 
brightest field stars near G174 and G177 have 
$J-K$ colors between 1 and 2, in broad agreement with what is seen in Figure 3.
However, the brightest stars have M$_K = -8.4$, 
which is roughly 0.3 mag fainter than measured from the CFHT data.

	The angular resolution of the CFHT data is almost a factor 
of two worse than the NICMOS data, and so the possibility that the difference 
in peak brightness is due to crowding is an obvious avenue for investigation. 
In an effort to determine if the difference in peak stellar brightness 
between the NICMOS and CFHT data is due to 
image quality, synthetic $H$ and $K$ datasets with stellar 
densities comparable to that in the G177 field were created using routines in 
the IRAF ARTDATA package; the G177 field was selected for detailed modelling 
because it has a stellar density that is similar to the CFHT bulge field. 

	Synthetic $H$ and $K$ datasets with FWHM = 0.19 
arcsec and FWHM = 0.35 arcsec were constructed. An AGB sequence 
peaking at K = 16 and with a power-law exponent -0.3, which is comparable 
to that seen in the Galactic bulge (e.g. Davidge 2000b), 
was created. The density of AGB stars was fixed according to 
the number of objects observed in the top $K$ magnitude interval in the 
G177 field. An RGB component was also included, with the peak brightness and 
relative number density of these stars, measured with respect to the AGB, 
based on those seen in the outer regions of M32 by Davidge (2000a). 
The AGB and RGB sequences were terminated at 
$K = 20$, which is 2.5 mag below the approximate faint limit of the CFHT 
data; over 10000 stars were thus added to each frame.
Finally, all of the stars were assigned an $H-K$ color of 0.0, and the 
PSF was assumed to be a gaussian. 

	The simulations are idealised in that they do not include the effects 
of, among other factors, (1) residuals in the flat-field and thermal background 
patterns, (2) anisoplanicity, and (3) intrinsic star-to-star color variations. 
Hence, the observed scatter in the $(K, H-K)$ CMDs obtained from the simulated 
images, which are shown in Figure 6, is significantly smaller than in the 
actual observations. Nevertheless, these simulations indicate that (1) crowding 
produces a modest population of spuriously bright sources in the G177 field 
even when FWHM = 0.19 arcsec, and (2) the number of such sources increases only 
by a factor of $1.5 - 2$ when the FWHM = 0.35 arcsec.

	The most conspicuous artifacts of crowding in 
Figure 6 are sources that are 0.4 - 0.5 magnitude brighter than 
the AGB-tip; while two such objects are seen in the 0.19 arcsec dataset, three 
are present in the 0.35 arcsec data. There is also a population of objects 
$0.05 - 0.15$ magnitude above the AGB-tip in both datasets. 
Aside from the easily identifiable population of 
blends, in both cases the AGB-tip at $K = 16$ is well-defined, 
although with 0.35 arcsec FWHM images the peak brightness might be infered to 
be 0.1 mag brighter than when the FWHM = 0.19 arcsec. Thus, 
crowding will cause the peak stellar brightness measured from the CFHT data to 
be at most 0.1 mag brighter than that measured from the NICMOS data.

	These simulations suggest that the difference in peak brightness with 
respect to the Stephens et al. (20001b) data is not due entirely to crowding. 
Perhaps the photometric calibrations of the two datasets differ 
at the 0.1 -- 0.2 magnitude level, and this speculation 
will require an independent set of observations to be confirmed. For the time 
being, it is worth noting that the peak brightness and number density of the 
brightest stars measured in the CFHT bulge field are consistent with 
the central M31 field (\S 4). Not only does this consistency lend 
confidence to the photometric calibration of the CFHT data, which were recorded 
over two observing runs, but it also reinforces the nature of the 
brightest objects as individual stars, and not artifacts of crowding.
The excellent agreement between the CFHT 
bulge field and BW LFs in Figure 4 is also reassuring. 

\subsection{Comparison with M32}

	The RGB-tip brightnesses of M31 and M32 agree to within 0.2 mag in $I$ 
(Davidge \& Jones 1992, Davidge 1993), suggesting that these galaxies are 
roughly equidistant, and this simplifies efforts to compare the bright stellar 
contents of these systems. Such a comparison is of interest since 
there are indications that M31 and M32 interacted in the past (Byrd 1975, 1978; 
Sofue \& Kato 1981), raising the possiblity of 
co-ordinated major star-forming episodes. 
Moreover, Luppino \& Tonry (1993) studied surface brightness flucuations 
in both galaxies at infrared wavelengths, and found significant 
differences in the characteristic flucuation brightness. 
Differences in the bright stellar contents of these systems might then 
be expected.

	The $K$ LFs of the M32 outer and M31 bulge fields are compared 
in Figure 7, where the M32 LF has been scaled 
to match the stellar density in the M31 bulge field using the $r$-band surface 
brightnesses measured by Kent (1987). It is evident that the M31 and M32 LFs 
differ by roughly a factor of three between $K = 16.5$ and $K = 17.5$, and 
the remainder of this section is devoted to exploring possible causes of this 
difference.

	The stellar density in the M31 bulge field is significantly higher than 
in the M32 outer field ($\mu_r = 18.8$ for the M31 bulge field versus $\mu_r = 
22.5$ for the outer M32 field), raising the possibility that the difference in 
Figure 7 could be due to crowding; indeed, the relative differences between the 
M31 bulge and M32 LFs in Figure 7 increases towards fainter brightnesses, 
as expected if crowding were a factor. However, simple arguements
suggest that crowding does not significantly affect star counts in the M31 
bulge field at the brightnesses where differences with respect to M32 are 
seen. If it is assumed that the M31 bulge 
and M32 have similar stellar contents, then the 
scaled M32 LF predicts that there will be 1850 stars with $K = 18 \pm 0.25$ 
in the M31 bulge field, most of which are evolving near the RGB-tip. If each 
resolution element has a diameter comparable to the FWHM of the PSF, then 
there will be roughly 200 blends among these sources in the bulge field, and 
these blended objects will appear as sources with $K \sim 17.5$; after 
correcting for incompleteness, this is roughly 10\% of the objects detected 
with $K = 17.5 \pm 0.25$ in the bulge field. A similar calculation shows that 
an even smaller fraction of the sources at $K \sim 17$ are blends. Thus, 
blending can not produce the factor of three difference between the M31 and M32 
datasets. 

	The comparison in Figure 7 assumes that M31 and M32 are equidistant. 
However, the observed differences could be produced if the bright stellar 
contents of these systems are in fact similar but their distance moduli differ 
by $\sim 0.5$ dex, in the sense that M32 is 
more distant. This is not consistent with 
the RGB-tip brightnesses of these systems. Davidge (1993) finds that 
the RGB-tip occurs at $I = 20.7$ in M31. While the RGB-tip is not well-defined 
in M32, it appears to be at least as bright as in M31, and may even be 
0.2 mag brighter (Freedman 1989; Davidge \& Jones 1992). If M32 is closer than 
M31 then the actual differences between the bright stellar contents of the two 
fields will be even greater than indicated in Figure 7.

	The comparison in Figure 7 assumes that the $r-K$ colors of the two 
fields are similar, and this may not be the case. Frogel et al. (1978) and 
Persson et al. (1980) published wide aperture ($d > 100$ arcsec) $V - K$ colors 
of M32 and M31, and these data indicate that the $V-K$ color of M32 is 
$\sim 0.1 - 0.2$ mag bluer than M31. However, such a difference in color 
can account for only part of the differences in Figure 7. 

	In summary, crowding and differences in broad-band colors may account 
for roughly one third of the difference seen in Figure 7; thus, it appears that 
the AGB LFs of the M31 bulge and M32 are intrinsically different, in that 
the relative density of AGB stars in M32 with $K$ between 
16.5 and 17.5 is lower than in the M31 bulge. In \S 4 this comparison is 
extended to data sampling the central regions of both galaxies.

\section{THE CENTRAL FIELD}

	The stellar density climbs rapidly with decreasing radius near the 
center of M31, with the result that the degree of crowding, which 
affects the completeness fraction and photometric errors, 
varies significantly across the central field. Given the steep density 
gradient, and also recognizing that changes in stellar content could occur 
over small angular scales near the galaxy center, it was decided to 
investigate the stellar content in 4 annuli centerered on the nuclear source P2.

	It is unlikely that individual stars are resolved with the current data 
within a few arcsec of the galaxy nucleus. Using near-infrared images of the 
Galactic Center that were processed to simulate the appearance of this field if 
viewed at the distance of M31, Davidge et al. (1997) concluded that any 
objects detected at $2\mu$m within $\sim 2$ arcsec of the M31 nucleus, which 
corresponds to the point at which the kinematic and photometric properties of 
M31 depart from the trends defined at larger radii (Kormendy \& Bender 1999), 
are likely blends of fainter stars, even 
when working at angular resolutions near the 
diffraction limit of a 4 metre telescope. However, as demonstrated below, the 
effects of blending decrease significantly at distances in excess of 2 arcsec 
of the nucleus; although blending still occurs at these radii, the effects can 
at least be monitored with simulations. Working outwards from 2 arcsec, 
the radial extent of each annulus was defined such that the number of stars 
between $K = 15.25$ and $K = 16.25$ was more-or-less evenly distributed 
between the annuli. The inner and outer radii of each annulus, with distances 
measured from the nuclear source P2, are listed in Table 1. 

	The $(K, H-K)$ CMDs of each annulus are shown in Figure 8; the $(K, 
J-K)$ CMDs are not considered here because of the relatively poor angular 
resolution of the $J$ data, and the resulting complications introduced by 
blending. The ridgelines of the AGB sequences in all 4 annuli and the bulge 
field are in excellent agreement. The brightest stars have $H-K$ 
= 0.4, which is consistent with the colors of late M giants in BW (e.g. 
Table 3 of Frogel \& Whitford 1987). The scatter predicted by the artificial 
star experiments matches the width of the giant sequences in all 4 annuli, 
indicating that random uncertainties in the photometric measurements, rather 
than an intrinsic dispersion in stellar properties, dominate the width of the 
CMDs.

	The CMDs of annuli 3 and 4 and the bulge field are very similar: in 
each case the main stellar sequence terminates near $K = 15.7 - 15.6$, and 
there is a spray of stars $\sim 0.1 - 0.2$ mag above this point. It is also 
worth noting that the bright portions of the CMDs of annuli 3 and 4 
are similar to the simulations shown in 
Figure 6. As for annuli 1 and 2, the AGB sequences 
are continuous to brighter values than in annuli 3 and 4, 
peaking near $K = 15.5$ in annulus 2, and $K = 15.3$ in annulus 1.

	An increase in peak stellar brightness towards progressively more 
crowded environments is a classic signature of blending. The stellar density 
in annulus 1 is roughly twice that in annulus 4, and so pairs of sub-regions 
in annulus 4 were summed to create simulated fields with densities comparable 
to that in annulus 1. Two pairs of $200 \times 200$ pixel 
regions in annulus 4 were summed, and the brightnesses of stars in the summed 
images were then measured using DAOPHOT. The CMDs of sources in the sub-regions 
prior to, and after, summing are compared in Figure 9. The CMDs of sources in 
the summed fields and in annulus 1 are very similar, lending confidence to the 
simulations. It is evident from Figure 9 that a factor of two increase in 
stellar density has a marked influence on the bright portions of the CMD at 
these densities, in that the peak brightness is elevated by 0.1 - 0.2 mag in 
$K$. Hence, the apparent trend of increasing peak giant branch brightness with 
decreasing radius in Figure 8 is the result of blending; this is contrary 
to the conclusion reached by Davidge et al. (1997), who did not have the 
benefit of data obtained at larger radii to gauge the effects of blending.

	The number density of infrared-bright stars scales with 
$r-$band surface brightness throughout the inner bulge of M31. This is 
demonstrated in Figure 10, where the $K$ LFs of annuli 1 -- 4 are compared with 
the $K$ LF of the bulge field, after the latter was scaled to match the 
mean $r$-band surface brightness in each annulus based on the Kent (1987) 
light profile; the LFs in this figure have been corrected for 
incompleteness, and are restricted to the bright end where sample 
incompleteness does not exceed 70\%. The LFs of annuli 3 and 4 are well 
matched by the LF of the bulge field at all brightnesses. There is an excess 
number of sources with $K = 15.5$ in annuli 1 and 2 when compared with the 
scaled bulge field population, and the simulations discussed 
in the previous paragraph suggest that this is the result of crowding, although 
the agreement at $K = 16$ and $K = 16.5$ is excellent. It thus 
appears that the radial distribution of the brightest stars at infrared 
wavelengths in the M31 bulge tracks the $r-$band light profile of the galaxy, 
even to within a few arcsec of the nucleus.

	Local mass density influences the star-forming history of regions 
within galaxies (e.g. Bell \& de Jong 2000; Martinelli, Matteucci, \& 
Colafrancesco 1998; Franx \& Illingworth 1990); galaxy-to-galaxy comparisons 
between areas having similar densities will remove this dependence, and thereby
provide a better means of searching for inherent differences in stellar content.
Davidge et al. (2000) obtained $H$ and $K$ observations with 0.12 
arcsec FWHM resolution of the central regions of M32, and the mean surface 
brightness in Region 3 of the Davidge et al. (2000) study is comparable to that 
in annuli 2 and 3 of the M31 central field. The $K$ LFs of M32 Region 3 and the 
sum of annuli 2 and 3 are compared in Figure 11. The agreement between the M31 
and M32 LFs is much better than in Figure 7, although the M31 and M32 LFs 
differ by a significant amount at $K = 16.5$, and the M31 LF falls consistently 
above that of M32 when $K > 16.5$, as would be expected from Figure 7. 
Nevertheless, the improved agreement between the 
M31 and M32 LFs at the faint end suggests 
that crowding may affect significantly the comparison in Figure 7 at $K = 
17.5$. We conclude that while the comparisons between the LFs of the M31 bulge 
and M32 in Figures 7 and 11 are {\it suggestive} of a difference in 
bright stellar content, in the sense that the number density of 
moderately bright AGB stars is lower in M32 than in the bulge of M31, the 
results are far from conclusive, and should be confirmed with data having 
higher angular resolutions to reduce the effects of crowding.

\section{DISCUSSION \& SUMMARY}

	Images with angular resolutions approaching the diffraction limit of 
the 3.6 metre CFHT have been used to investigate the infrared-bright stellar 
content of the inner bulge of M31. The number density of the 
brightest stars in $K$ scales with $r-$band surface 
brightness in the central few arcmin of the galaxy. This 
indicates that (1) these objects are well mixed throughout the main body of the 
inner bulge, and thus belong to a population that formed in a highly 
coherent manner, and (2) the  brightest stars do not belong 
to the disk (see also Rich et al. 1993), since the light profile of the galaxy 
is dominated by the bulge at small radii. Previous investigations of the 
stellar content of the M31 bulge are reviewed in \S 5.1, and it is concluded 
that the bulge is dominated by a population of old stars. Building on this 
result, it is argued in \S 5.2 that the infrared-bright AGB 
stars are old objects, and this interpretation is consistent with 
the spatial distribution of these stars.

\subsection{The Stellar Content of the M31 Bulge}

	In recent years there has been a growing realization 
that the bulges of spiral galaxies do 
not evolve in isolation; rather, the diversity evident in the central 
morphological (e.g. Carollo, Stiavelli, \& Mack 1998) and overall structural 
(Carollo 1999) characteristics of spiral galaxies suggests that bulges are 
influenced by external environmental factors. The disk is an obvious source of 
star-forming material, and evidence that the evolution of the bulge and 
surrounding disk are coupled comes from the correlated ages 
(Peletier \& Balcells 1996) and structural 
characteristics (Andredakis, Peletier, \& Balcells 1995) of these systems. 
Bar instabilities (e.g. Friedli \& Benz 1993, 1995), galaxy-galaxy 
interactions (e.g. Barnes \& Hernquist 1992, Mihos \& Hernquist 1996), and 
dynamical friction (Noguchi 1999, 2000) are three mechanisms by which gas 
and stars from the disk can be channeled into the central regions of bulges.

	Is there evidence for a mixture of stellar ages in the bulge of M31?
Imaging surveys indicate that the bulge of M31 does not contain 
young stars. The brightest members of a very 
young population have blue colors, and 
Brown et al. (1998) argue that the majority of resolved 
sources detected in the UV near the nucleus of M31 are evolving 
on the extended HB rather than the main sequence. Based on the 
faint limit of their data, Brown et al. (1998) conclude that the youngest 
UV-bright stars in the M31 bulge have an age in excess of 250 Myr. 

	Spectroscopic studies at visible wavelengths suggest that stars 
spanning a range of ages are present within a few arcsec of the center of M31. 
Both Davidge (1997) and Sil'Chenko et al. (1998) find evidence for an age 
gradient near the center of the galaxy, indicating that any intermediate age 
component in the central regions of M31 is centrally concentrated, and has a 
spatial distribution that differs from that of the main body of the bulge. 
The intermediate-age population does not dominate the innermost regions of the 
M31 bulge. Bica et al. (1990) used evolutionary synthesis techniques to 
investigate the integrated visible spectrum of the central few arcsec of M31 
and found that, even near the galaxy center, the bulge is dominated by an old 
population. In addition to an intermediate-age component that contributes a 
modest fraction of the integrated light at visible wavelengths, Bica et al. 
(1990) also found evidence of stars as young as 10 Myr, although the 
presence of such a population is not supported by UV imaging data. 

	There is other evidence indicating that 
the bulges of type Sbc and earlier spirals are 
dominated by old populations. In the case of the 
Milky-Way, Feltzing \& Gilmore (2000) examined HST images of Galactic bulge 
fields, and concluded that disk contamination is responsible for the relatively 
bright blue stars that have heretofore been associated with an intermediate-age 
population. Once disk contamination is taken into account, Feltzing \& 
Gilmore (2000) conclude that the main body of the bulge has an old age, in 
agreement with the bulge (Minniti 1995) globular cluster system (e.g. 
Ortolani et al. 1995; Fullton et al. 1995). Finally, the colors of nearby 
(Peletier et al. 1999) and distant (Abraham et al. 1999) spiral galaxies 
indicate that the bulges of these systems contain a significant component 
that formed early-on. 

	The central regions of the Galaxy contain stars spanning a range 
of ages. There are compact star clusters near the GC (Cotera et al. 1996, 
Figer et al. 1999) with ages of a few Myr, indicating that star formation can 
(and does) occur in the innermost regions of the Galaxy. While it can be 
argued that these clusters do not belong to the bulge, 
they have short dynamical timescales (e.g. Figer et 
al. 1999), and hence quickly evaporate and contribute to the field population 
in the innermost regions of the galaxy. We speculate that the centrally 
concentrated intermediate age population in M31, which has a spatial 
distribution that is different from the underlying bulge, may be an artifact of 
an earlier star-forming event similar to that recently experienced by the GC.

	In summary, while signatures of an intermediate age 
population may be present in integrated spectra of the center of M31, 
this population is centrally concentrated, and is not a major constituent of 
the bulge, which is dominated by old stars. 
A dominant old population is consistent with 
the non-solar [Mg/Fe] ratio of stars in the M31 bulge found by Davidge (1997), 
which is suggestive of a rapid initial chemical enrichment by SN II, 
as expected if the M31 bulge experienced a rapid (t $< 1$ Gyr) collapse 
during early epochs. The bulge of M31 has apparently not been 
subject to the external sources of star forming material or intermediate age 
and younger stars discussed in the opening paragraph of this section. 

\subsection{The Nature of the Brightest AGB stars in the M31 Bulge}

	The brightest AGB stars in the M31 bulge follow the integrated light 
profile of the galaxy and have a peak brightness that does not 
change measureably with radius. The brightest stars at infrared wavelengths 
are therefore well mixed throughout the inner bulge; these stars can not belong 
to an intermediate-age population, which the spectroscopic 
data suggests is more centrally concentrated than 
the main body of the bulge (Davidge 1997). Thus, the infrared-bright AGB stars 
likely belong to the old population that dominates the bulge of M31. 

	The brightest stars in the compact elliptical galaxy M32 share 
common characteristics with their counterparts in the bulge of M31 in that they 
(1) have a similar peak brightness, and (2) are uniformly mixed throughout the 
galaxy (Davidge 2000a; Davidge et al. 2000).
The brightness of the AGB-tip is more sensitive to metallicity than age 
in stellar systems with ages exceeding a few Gyr. 
If the MDF of the M31 bulge field is similar to that in BW, then the 
majority of stars in this field will have metallicities close to solar, and 
there will also be a population of super metal-rich objects (McWilliam \& Rich 
1994). AGB-tip stars in old moderately metal-rich globular clusters have 
brightnesses approaching those of the brightest stars in M32 and the M31 bulge; 
for example, stars as bright as M$_K = -8.5$ are present in the [Fe/H] = -0.34 
(Harris 1996) globular cluster NGC 6553 (Guarnieri, Renzini, \& Ortolani 1997), 
and even brighter AGB-tip stars should occur in a population that is more 
metal-rich. We thus suggest that the brightest AGB stars in M32 and the 
Galactic and M31 bulges are very metal-rich: these stars are bright 
not because of their age, but because of their chemical 
composition. Given that the mean metallicity of M32 is lower than that in the 
bulge of M31 (Bica et al. 1990), then the number density of bright AGB stars 
at a given surface brightness should be lower in the former than in the 
latter, and this is consistent with the comparisons between the $K$ LFs 
of these systems, which are discussed in \S 3 and 4, although clearly the 
differences seen between the bright stellar contents of M32 and the M31 bulge 
need to be confirmed with data having higher angular resolutions.

	The brightest stars in M32 and the M31 bulge have similar peak 
brightnesses; given that the brightness 
of the AGB-tip depends more on metallicity than age among old populations, 
then it might be anticipated that the brightest stars in these systems have 
similar metallicities, and this prediction can be tested spectroscopically. 
The targets are relatively bright, with $K = 15.5 - 16.0$; hence, 
the observational challenge is not one of obtaining a large 
S/N ratio, but of resolving individual stars in crowded 
environments. AO-fed integral-field spectrographs will be 
essential for obtaining uncontaminated spectra of these stars.

\vspace{0.3cm}
	It is a pleasure to thank the anonymous referee for providing a 
comprehensive report containing suggestions that greatly improved the paper.

\clearpage

\begin{table*}
\begin{center}
\begin{tabular}{cc}
\tableline\tableline
Annulus & Radii \\
 \# & (arcsec) \\
\hline
1 & 2.0 -- 7.8 \\
2 & 7.8 -- 11.9 \\
3 & 11.9 -- 15.3 \\
4 & 15.3 -- 24.0 \\
\tableline
\end{tabular}
\end{center}
\caption{Radial Intervals for Annuli in the Central Field}
\end{table*}

\clearpage

\begin{center}
FIGURE CAPTIONS
\end{center}

\figcaption
[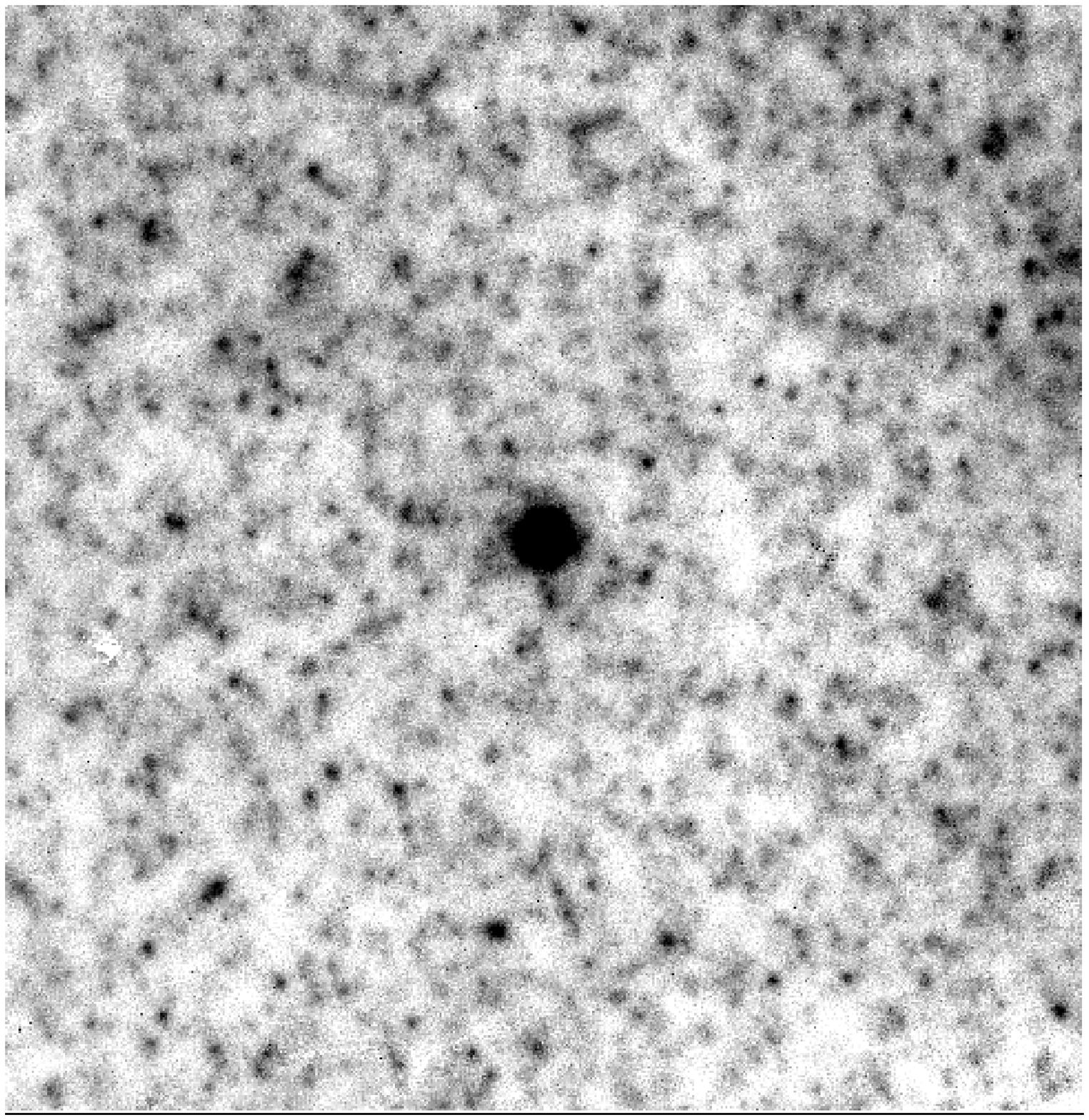]
{The final $Ks$ image of the bulge field. North is at the top, and 
east is to the left. The image covers $34 \times 34$ arcsec, and the angular 
resolution is 0.35 arcsec FWHM. The bright central source is the star GSC 
02801--02015, which served as the reference source for AO compensation.}

\figcaption
[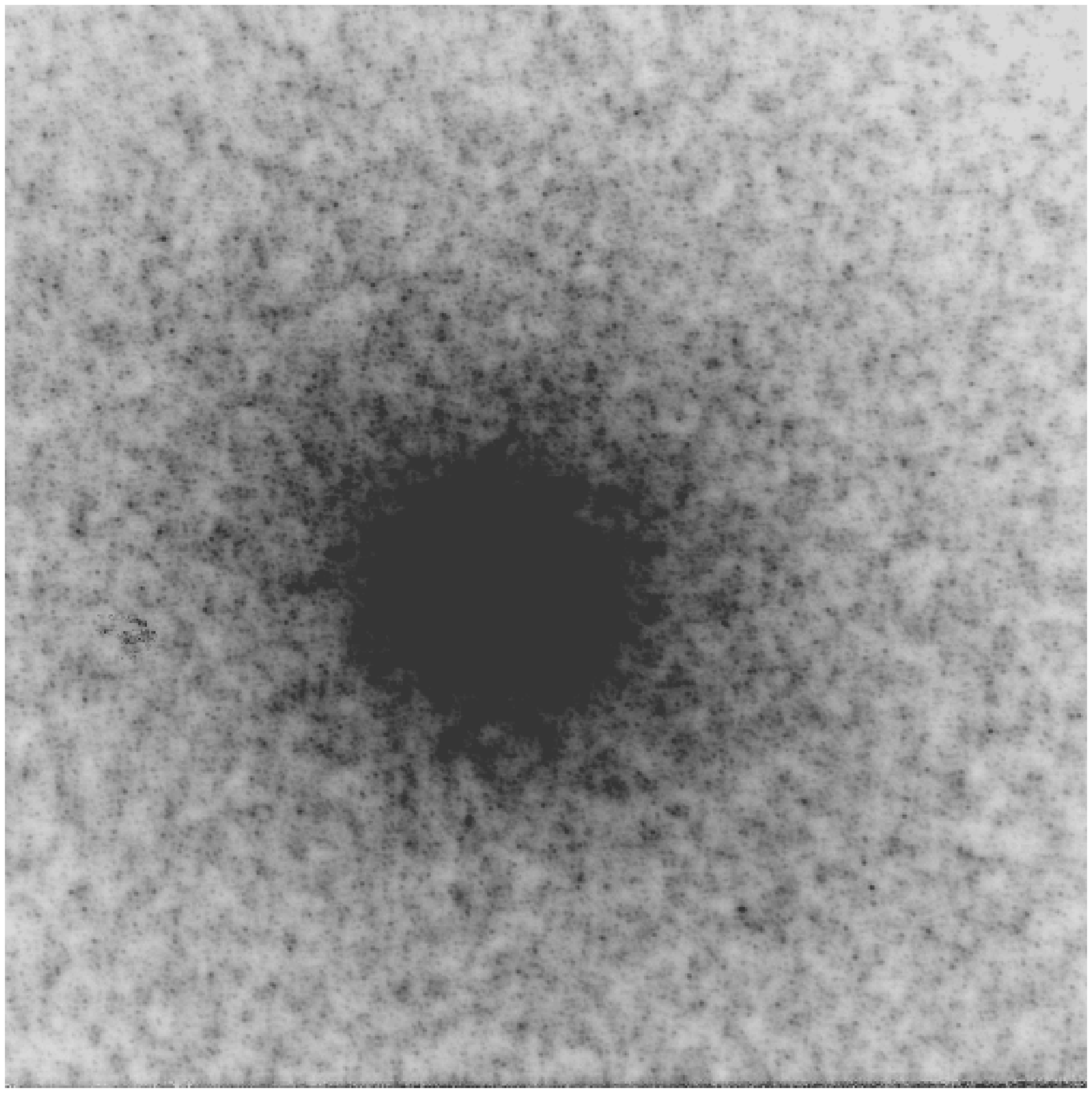]
{The final $Ks$ image of the central M31 field. North is at the top, 
and east is to the left. The image covers $34 \times 34$ arcsec, and the 
angular resolution is 0.17 arcsec FWHM.}

\figcaption
[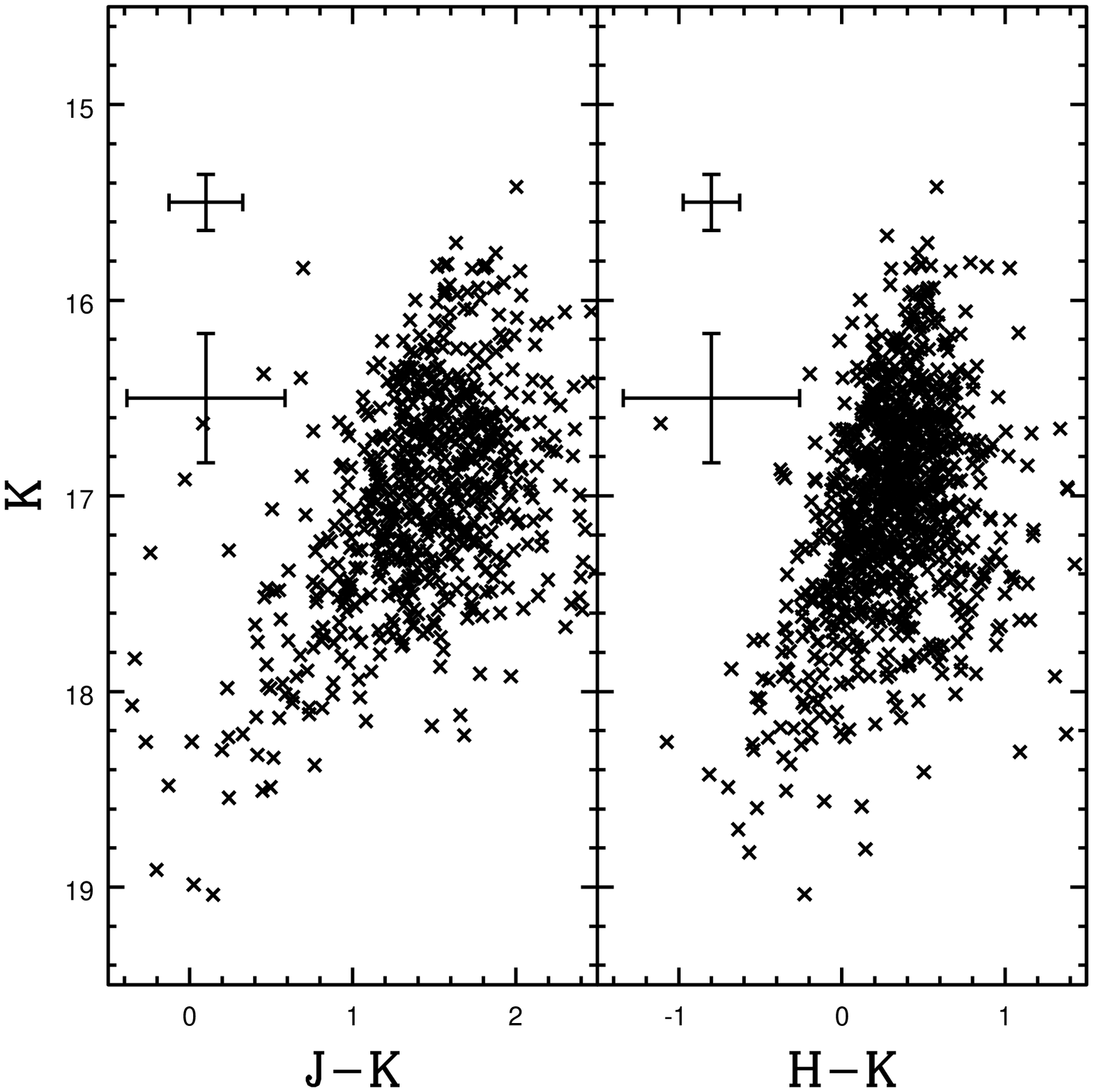]
{The $(K, J-K)$ and $(K, H-K)$ CMDs of the M31 bulge field. The 
error bars show the uncertainties predicted from the artificial star 
experiments, which assume a constant PSF for each field. The good 
agreement between the predicted and observed scatter indicates that 
anisoplanicity does not contribute significantly to the photometric errors.}

\figcaption
[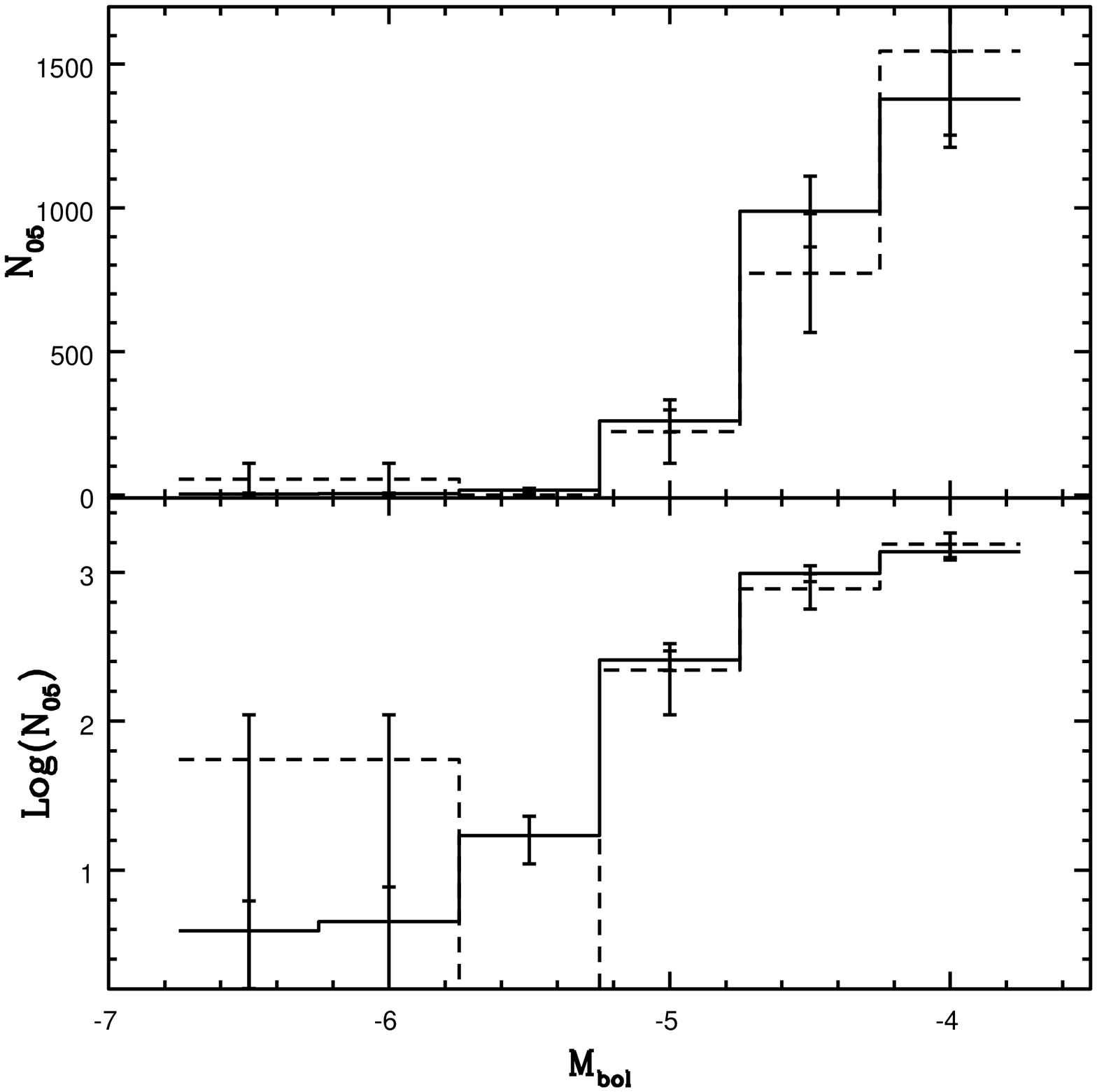]
{The bolometric LFs of the M31 bulge field (solid line) and BW (dashed line). 
The BW LF was constructed from the bolometric magnitudes listed in Table 1 of 
Frogel \& Whitford (1987), but assuming a distance modulus 
of 14.5. $N_{05}$ is the number of stars per 0.5 mag bin corrected for 
incompleteness. The error bars in the M31 LF reflect the uncertainties 
introduced by counting statistics and the completeness corrections, while 
the error bars for the BW LF include only counting statistics. The BW LF has 
been scaled to match the number of stars in the M31 LF when M$_{bol} 
< -4$. Note that the two LFs agree to within the estimated uncertainties, 
indicating that the bright stellar content of the M31 bulge is similar 
to that in BW.}

\figcaption
[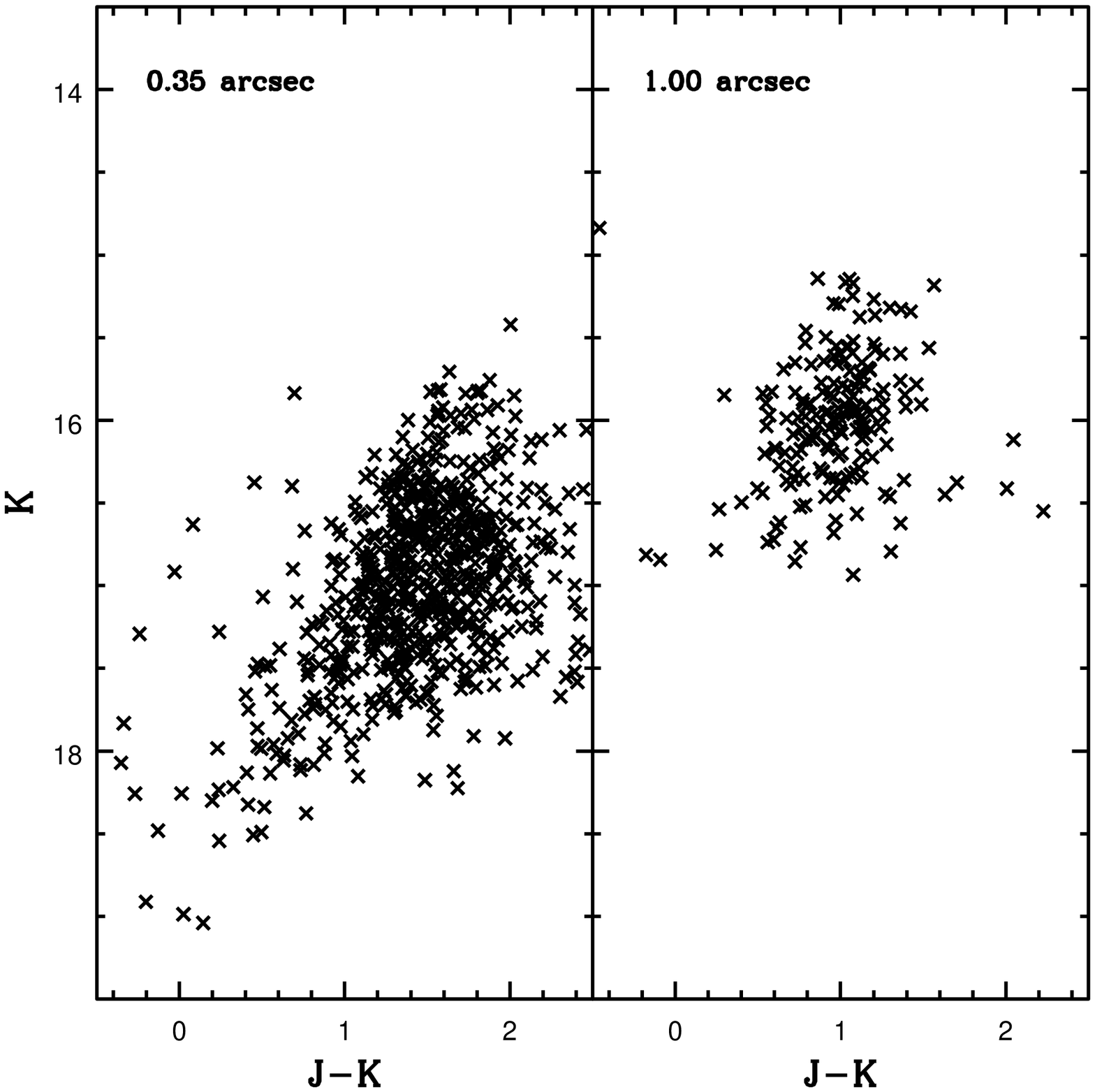]
{The $(K, J-K)$ CMD of the M31 bulge field, as observed with 
0.35 arcsec FWHM image quality, compared with the CMD of the same field after 
convolving with a Gaussian to simulate 1 arcsec FWHM image quality. Note 
the difference in peak stellar brightness.}

\figcaption
[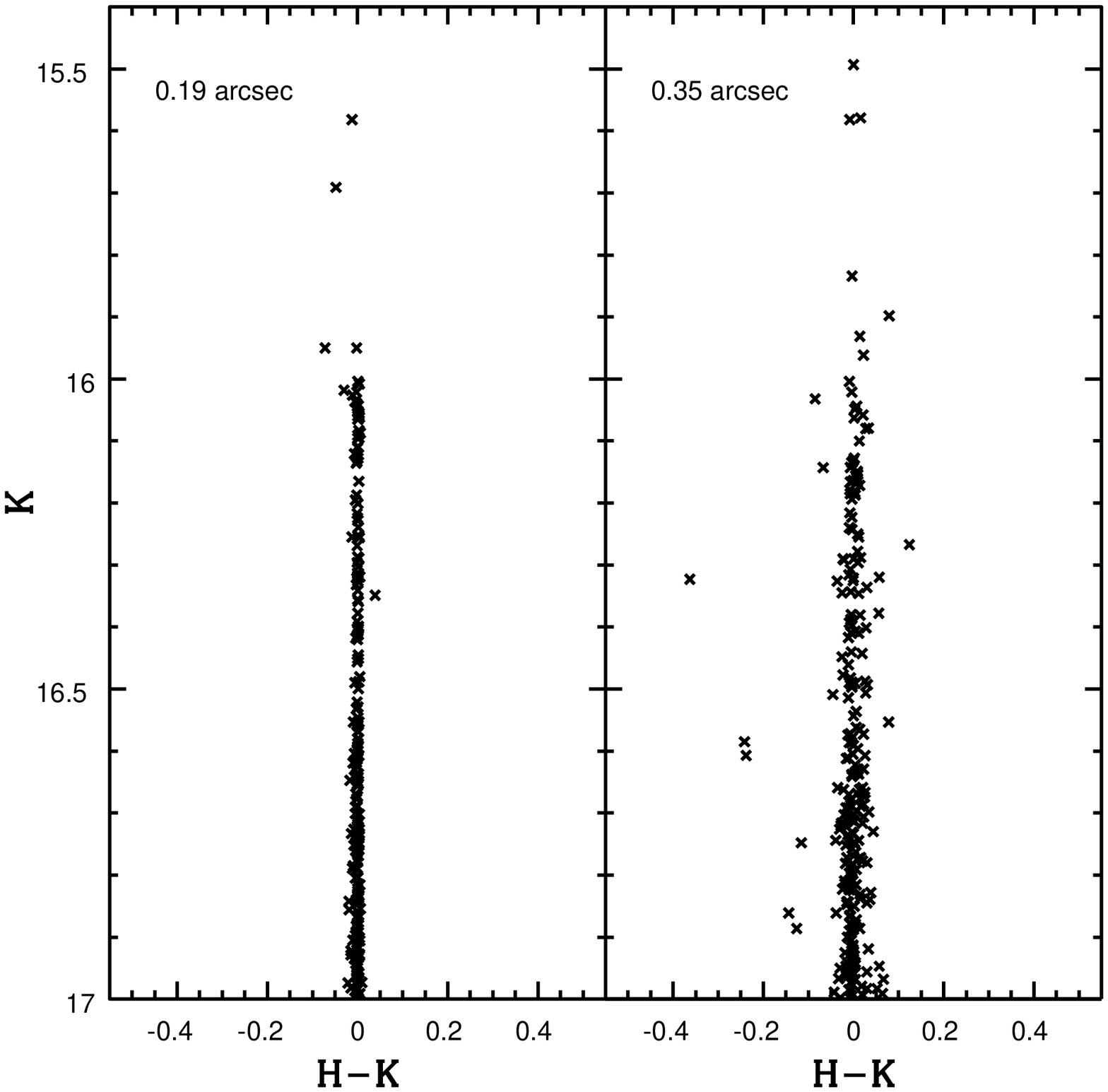]
{The $(K, H-K)$ CMDs obtained from simulated images of a field with a stellar 
density identical to that in the Stephens et al. (2001a) G177 field. The 
simulated datasets have FWHM = 0.19 arcsec, in agreement with the Stephens et 
al. data, and 0.35 arcsec, which is the angular resolution of the 
CFHT bulge field data. Additional details of the models can be found in the 
text. Note that in both cases there is a smattering of blended objects above 
the AGB-tip, which occurs at $K = 16$; the brightest of these are well 
separated from the main body of stars in the CMD, and hence are easily 
identified. The simulations indicate that when FWHM = 0.35 arcsec 
the brightness of the AGB-tip may be overestimated by at most 0.1 magnitude 
in $K$ when compared with data having FWHM = 0.19 arcsec.}

\figcaption
[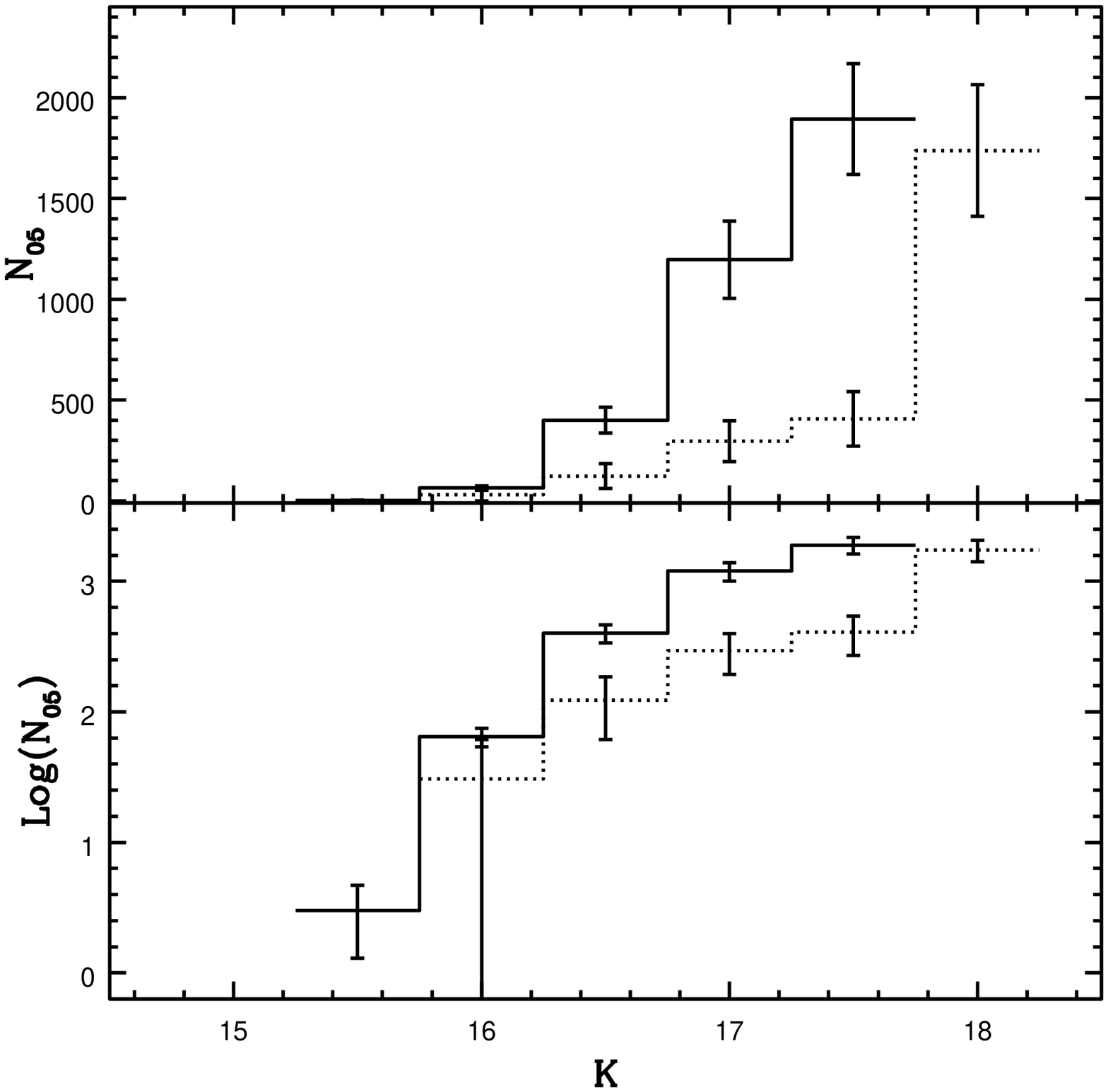]
{The $K$ LF of the M31 bulge field (solid line), based on sources 
detected in both $H$ and $K$, compared with the $K$ 
LF of the M32 outer field (dotted line) observed by Davidge (2000a). N$_{05}$ 
is the number of stars per 0.5 mag interval. The 
LFs have been corrected for incompleteness using results 
from artificial star experiments. The M32 LF 
has been scaled to match the $r-$band surface brightness of the M31 field using 
the measurements made by Kent (1987). The error bars show the uncertainties 
due to counting statistics and the completeness corrections. Note the 
disagreement between the M32 and M31 LFs between $K = 16.5$ and 17.5.}

\figcaption
[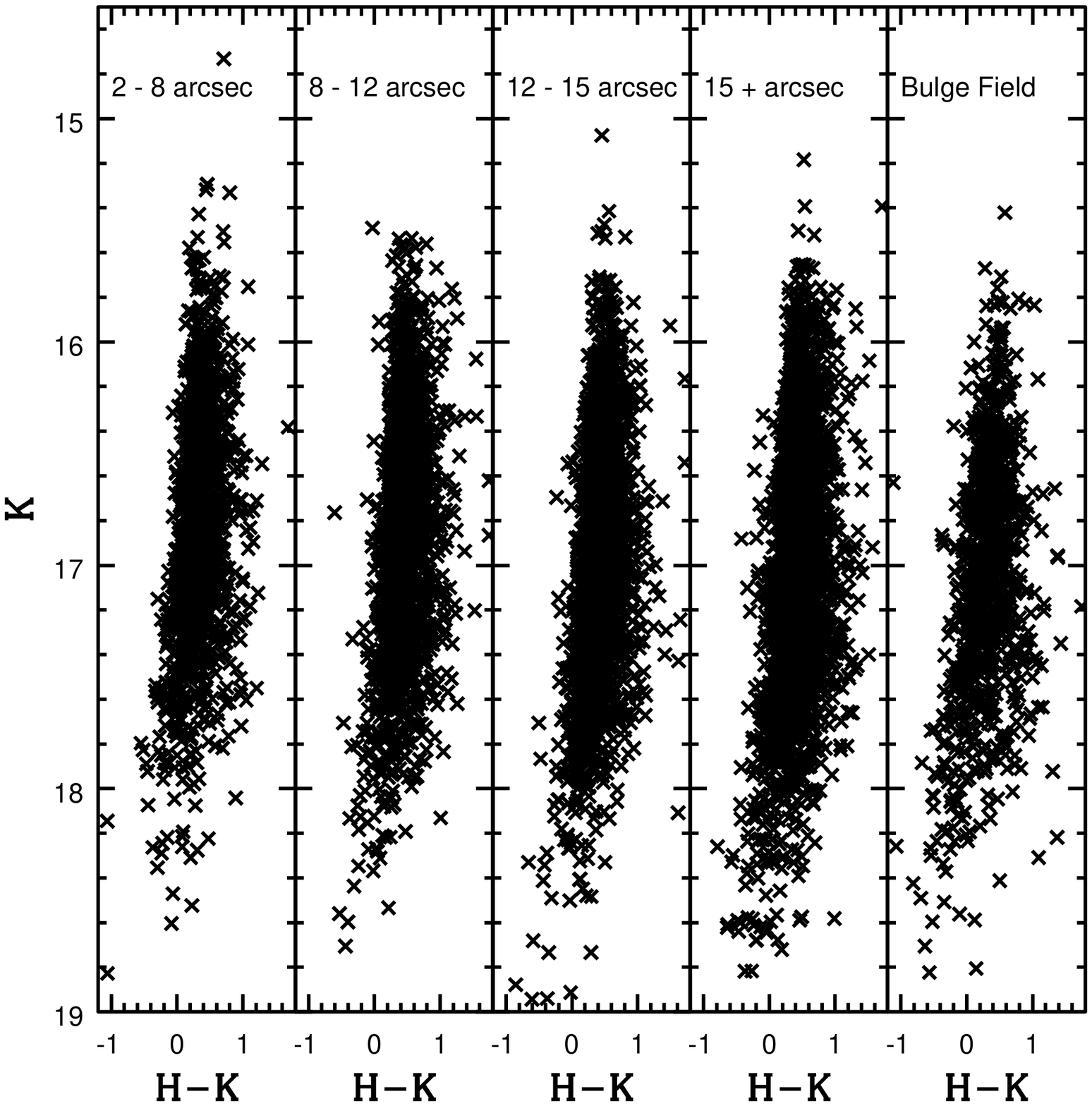]
{The $(K, H-K)$ CMDs of stars at various distances from the center of M31, 
including the bulge field. The scatter in the data is well matched by the 
photometric errors predicted by the artificial star experiments.}

\figcaption
[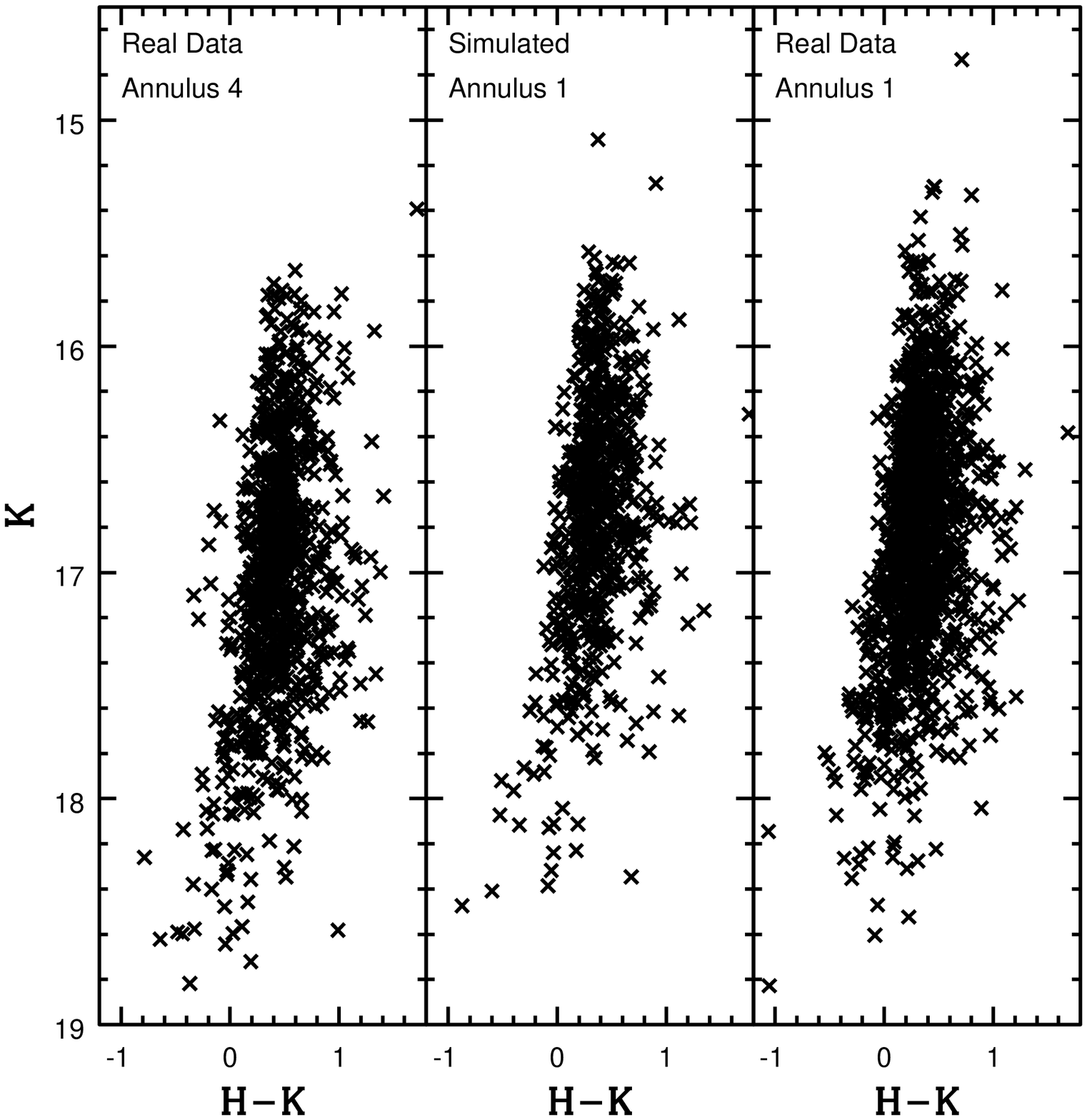]
{The effects of image blending are investigated in this figure. The left hand 
panel shows the composite $(K, H-K)$ CMD of four 200 $\times$ 200 pixel 
sub-regions in annulus 4. The middle panel shows the composite $(K, H-K)$ CMD 
after these sub-regions were paired and summed to simulate the stellar density 
in annulus 1. These simulations indicate that a factor of two increase in 
stellar density elevates the peak stellar brightness by 0.1 -- 0.2 
mag near the nucleus of M31. The CMD constructed from the summed dataset is 
remarkably similar to the $(K, H-K)$ CMD of annulus 1, which is shown in the 
right hand panel.}

\figcaption
[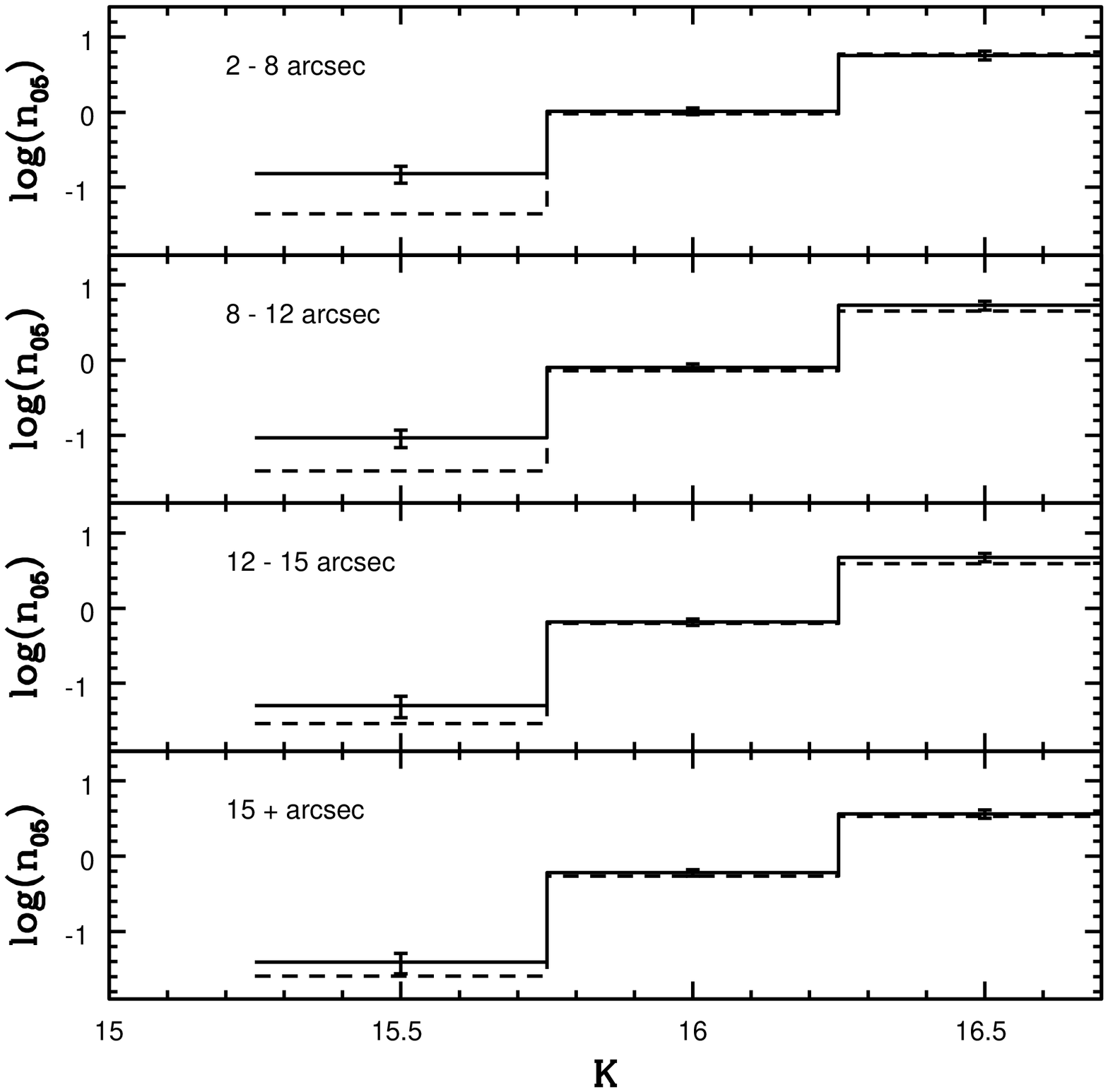]
{The $K$ LFs of the 4 annuli in the central field (solid lines), 
constructed from stars detected in both $H$ and $K$, 
compared with the $K$ LF of the bulge field (dashed line), which has been 
scaled to match the $r-$band surface brightness in each annulus using the 
measurements published by Kent (1987). The LFs have been 
corrected for incompleteness, based on results from 
artificial star experiments. n$_{05}$ is the number of stars per 
square arcsec per 0.5 mag, and the error bars show the uncertainties 
introduced by counting statistics and the completeness corrections. 
Note the excellent agreement with the bulge field LF at $K = 16$ and 
$K = 16.5$ for each annulus. The tendency for the number density of sources 
at $K = 15.5 \pm 0.25$ in the central field to increase towards smaller radii 
at a rate that is faster than expected from the bulge field 
data is likely the result of blending.}

\figcaption
[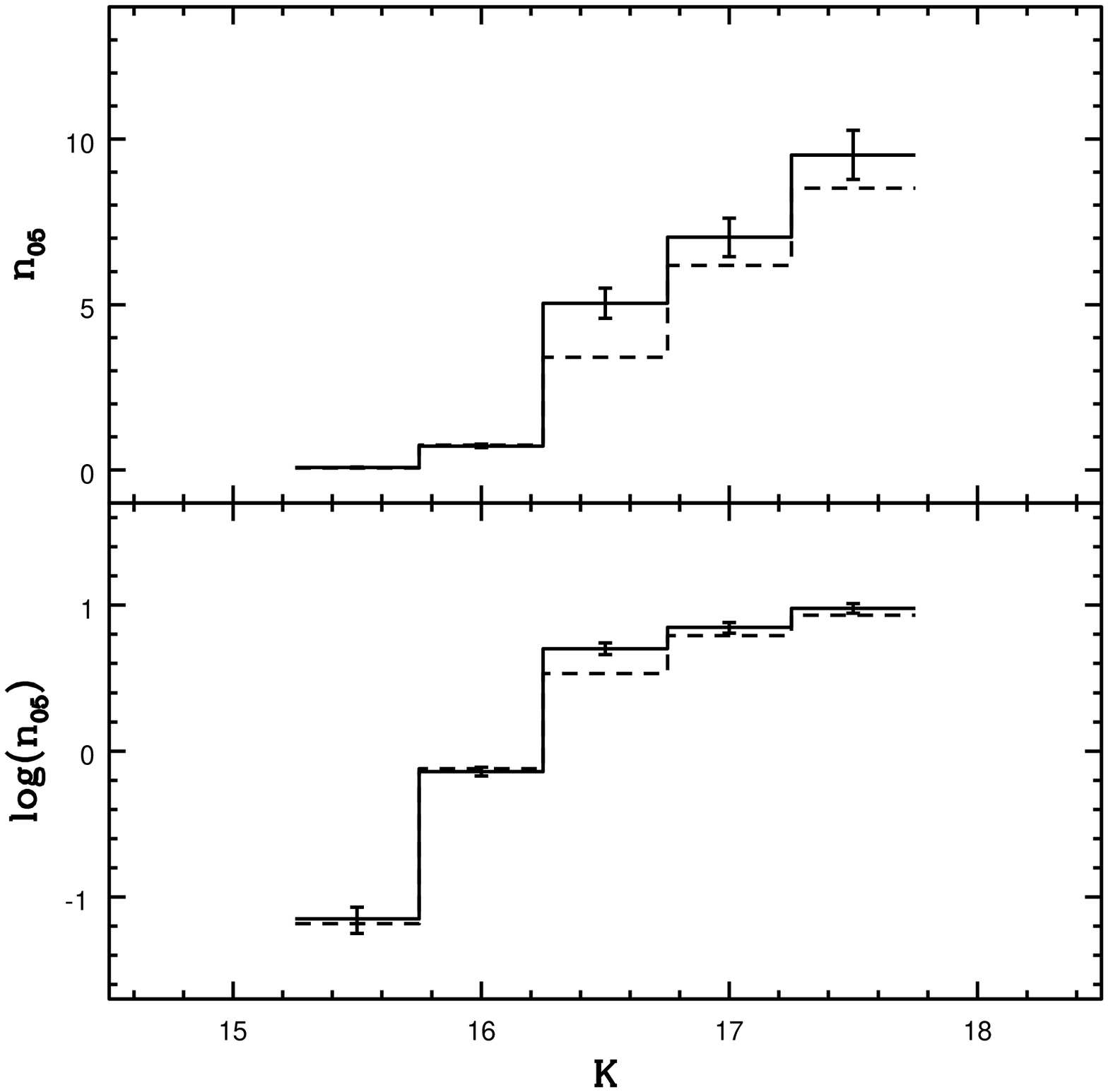]
{The $K$ LF of annuli 2 and 3 (solid line) compared with the 
LF of M32 Region 3 from Davidge et al. (2000), which has an $r-$band surface 
brightness comparable to that of M31 annuli 2 and 3. n$_{05}$ is the number of 
stars per square arcsec per 0.5 mag interval. The LFs have been corrected 
for incompleteness based on the results from artificial star 
experiments. Note that the M31 and M32 LFs differ significantly 
only when $K = 16.5$, although the M31 LF falls consistently above that of 
M32 when $K > 16.5$, as expected from Figure 7.}

\end{document}